\documentclass[
    ,final            
  ]
  {aipproc}

\usepackage{amssymb}  
\usepackage{graphicx,subfigure}
\usepackage{exscale}
\usepackage{textcomp}
\usepackage{enumerate}

\usepackage{color}

\usepackage[normalem]{ulem}  

\newcommand{\be}{\begin{equation}}
\newcommand{\ee}{\end{equation}}
\newcommand{\bea}{\begin{eqnarray}}
\newcommand{\eea}{\end{eqnarray}}

\layoutstyle{8x11single}

\begin{document}

\title{Baryon onset in a magnetic field}


\author{Alexander Haber}{
  address={Institut f\"{u}r Theoretische Physik, Technische Universit\"{a}t Wien, 1040 Vienna, Austria}
}

\author{Florian Preis}{
  address={Institut f\"{u}r Theoretische Physik, Technische Universit\"{a}t Wien, 1040 Vienna, Austria}
}

\author{Andreas Schmitt}{
  address={Institut f\"{u}r Theoretische Physik, Technische Universit\"{a}t Wien, 1040 Vienna, Austria}
}

\begin{abstract}

The critical baryon chemical potential for the onset of nuclear matter is a function of the vacuum mass and the binding energy. Both quantities are 
affected by an external magnetic field. We show within two relativistic mean-field models -- including magnetic catalysis, but omitting the anomalous magnetic 
moment -- that a magnetic field increases both the vacuum mass and the binding energy. For sufficiently large magnetic fields, the effect on the vacuum mass dominates and 
as a result the critical baryon chemical potential is increased.

\end{abstract}

\maketitle

\section{Introduction}

The phase structure of Quantum Chromodynamics (QCD) is changed by a background magnetic field $B$, as we know from the transition temperature of the crossover between 
the chirally broken and the (approximately) chirally symmetric phases, which decreases with $B$ \cite{Bali:2011qj}. We may thus, more generally, ask 
how the phase diagram in the plane of temperature and baryon chemical potential $\mu$ changes in the presence of a magnetic field. 
However, at non-vanishing $\mu$, the phase structure 
is not well understood even for $B=0$. One of the few features of the QCD phase diagram which is known with certainty is the liquid-gas phase transition, 
which, at zero temperature, is a first-order onset of nuclear matter. Here we address the question how the critical chemical potential of this onset is changed in the 
presence of a magnetic field. 

To understand the role of the magnetic field, we recall that the critical chemical potential for the onset is the energy needed to add a baryon to 
the system. In a non-interacting system, this energy would be identical to the vacuum mass of a baryon. In real-world nuclear matter, interactions reduce 
this energy, such that the critical chemical potential is the vacuum mass minus the binding energy. Therefore, we have to ask how a magnetic field 
acts, firstly, on the vacuum mass and, secondly, how it affects the binding energy. Regarding the vacuum mass, we may again split the effect of the magnetic field into 
two contributions, its effect on the quark masses and on the interactions between the quarks. 
It is known that a magnetic field tends to increase the chiral condensate and thus the 
quark masses \cite{Klimenko:1990rh,Gusynin:1994re}. This effect is called magnetic catalysis \cite{Shovkovy:2012zn} and has been observed for zero-temperature 
QCD on the lattice \cite{Bali:2011qj,D'Elia:2010nq} and in chiral perturbation theory \cite{Shushpanov:1997sf,Agasian:1999sx}. The main conclusion of the present 
calculation -- which is based on our work in Ref.\ \cite{Haber:2014ula} -- is that 
this effect, at least qualitatively, can be accounted for in simple relativistic mean-field models of nuclear matter.
The effect of the magnetic field on the interactions of the quarks is more difficult to incorporate. At least part of this effect may be accounted 
for by an effective description in terms of the anomalous magnetic moment. This has been done previously in the literature, but without
taking into account the vacuum contribution that is responsible for magnetic 
catalysis \cite{2000ApJ...537..351B,Broderick:2001qw,Sinha:2010fm,Rabhi:2011ej,Dexheimer:2011pz,Preis:2011sp,Dong:2013hta,deLima:2013dda,Casali:2013jka}. 
In future studies, it will be important to include both effects, magnetic catalysis and the anomalous magnetic moment, 
especially since the effect of the anomalous magnetic moment can be expected to counteract the result observed here \cite{Preis:2010cq,Andreichikov:2013pga}. 
Thus, our work includes the effect of the magnetic field on the vacuum mass for the first time, but certainly not in a complete way. The effect 
on the binding energy -- which results from the interactions between the {\it nucleons} -- {\it is} included, 
at least in the simplified version of relativistic mean-field models, whose parameters are fitted to the saturation properties of nuclear matter 
in the absence of a magnetic field. 

A more formal way of explaining our calculation is as follows. In phenomenological models of dense nuclear matter, usually the vacuum contribution is omitted, and one
works solely with the medium contribution. This ``no-sea approximation'' is sometimes justified with the phenomenological nature of the model, where the parameters are
fitted to reproduce properties of nuclear matter {\it in the absence} of the vacuum terms. This may seem unsatisfactory from a purely theoretical point of view,
and moreover is clearly not a controlled approximation of the full model. However, it has been shown that the contribution of the vacuum terms is indeed small if one is 
interested for instance in the equation of state of dense nuclear matter \cite{1988PhLB..208..335G,1989NuPhA.493..521G}. The purpose of the present work is to 
point out that a qualitative mistake is made if the ``no-sea approximation'' is naively applied for systems in a background magnetic field. The reason is that 
magnetic catalysis is a vacuum effect. In models where the degrees of freedom are quarks, such as the Nambu--Jona-Lasinio model, it is well-known that magnetic catalysis
is obtained from the contribution of the Dirac sea. Therefore, throwing out this contribution in phenomenological models for magnetized nuclear matter is equivalent
to ignoring the effect of magnetic catalysis. 

Besides the theoretical motivation in the context of the QCD phase diagram, our study may also be relevant for the phenomenology of compact stars. Compact stars are 
known to have huge magnetic fields, up to $10^{15}\, {\rm G}$ at the surface. In the interior, it is consistent with the stability of the star to expect magnetic fields 
that are several orders of magnitude larger, $10^{18-20}\,{\rm G}$, depending on the matter composition of the star \cite{Lai,Ferrer:2010wz}. In that case, 
dense matter in the core of the star will be affected on the QCD scale. Since we are only concerned with the baryon onset, our results are not directly applicable to 
the large densities in the interior of compact stars. Nevertheless, our calculation shows how to incorporate magnetic catalysis in dense nuclear matter, and it remains
to be seen whether it has a sizable effect on the equation of state and thus on observable quantities such as the mass and the radius of the star. Additionally, it is 
conceivable that huge magnetic fields are created in mergers of compact stars via a magneto-rotational instability \cite{Siegel:2013nrw}. 
Since direct observation of gravitational waves from such merger processes 
may yield information about the equation of state \cite{Read:2013zra}, additional data about strongly interacting matter in a magnetic field may become available in 
the future.

\section{Dirac sea contribution}

The free energy density of non-interacting, spin-$\frac{1}{2}$ fermions with mass $M$ and electric charge $q$ in a homogeneous magnetic field 
in $z$-direction with field strength $B$  is
\be \label{Omega}
\Omega = - \frac{|qB|}{2\pi} \sum_{\nu=0}^\infty \alpha_\nu\int_{-\infty}^\infty\frac{dk_z}{2\pi}\left[\epsilon_{k,\nu}
+T\ln\left(1+e^{-\frac{\epsilon_{k,\nu}- \mu}{T}}\right)+T\ln\left(1+e^{-\frac{\epsilon_{k,\nu}+ \mu}{T}}\right)\right] \, ,
\ee
where $T$ is the temperature, $\mu$ the chemical potential, $\alpha_\nu \equiv 2-\delta_{\nu 0}$, and 
\be \label{epskn}
\epsilon_{k,\nu} = \sqrt{k_z^2 + 2\nu|qB| + M^2} 
\ee
are the single-fermion excitations. 
In the mean-field models used below, all effects of the interactions can be absorbed into an effective mass and an effective chemical potential, such that the 
free energy is formally identical to that of non-interacting fermions. The important difference is that mass and effective chemical potential are determined 
self-consistently, i.e., they depend on $B$, $\mu$ and $T$. The free energy (\ref{Omega}) has three contributions, a Dirac sea contribution, and two medium contributions 
from fermions and anti-fermions (the latter being irrelevant for our zero-temperature calculation). The Dirac sea contribution is divergent and can, after 
regularization via the proper time method be written as
\be \label{sea}
\frac{B^2}{2} - \frac{|qB|}{2\pi} \sum_{\nu=0}^\infty \alpha_\nu\int_{-\infty}^\infty\frac{dk_z}{2\pi}\epsilon_{k,\nu} = \frac{B_r^2}{2} + \Omega_0
 - \frac{|qB|^2}{24\pi^2}\ln\frac{2|qB|}{\ell^2A^{12}} - \frac{|qB|^2}{2\pi^2}\left[\frac{x^2}{4}(3-2\ln x) 
+\frac{x}{2}\left(\ln\frac{x}{2\pi}-1\right)+\psi^{(-2)}(x)\right] \, , 
\ee
where we have included the free field contribution proportional to $B^2$, where $A$ is the Glaisher constant and $\psi^{(n)}$ the $n$-th polygamma function. 
We have separated the (divergent) $B=0$ contribution $\Omega_0$,
defined $x\equiv M^2/(2|qB|)$, and introduced the renormalization scale $\ell$ and the renormalized magnetic field via $B^2 = Z_q B_r^2$ with
$Z_q = 1+q_r^2/(12\pi^2)[\gamma+\ln(\ell^2/\Lambda^2)]$. Here, $\Lambda$ is an ultraviolet cutoff, $\gamma$ the Euler-Mascheroni constant, and 
$q_r$ the renormalized electric charge, such that $qB=q_rB_r$. For our purpose, only
the last contribution of Eq.\ (\ref{sea}), which depends on $M$ {\it and} $qB$, is relevant. We shall neglect the $B=0$ contribution $\Omega_0$, assuming
that its contribution to our final result is small \cite{Haber:2014ula}.   

\begin{table}[t]
\begin{tabular}{c|c|c}
\hline
 & Walecka & eLSM  \\ \hline\hline 
 (isospin-symmetric) nucleons & $N=n,p$ & $N;\, N^*\to N(1535)$ \\ \hline
 meson fields &  $\;\;$$\omega \to \omega(782) \, , \quad  \sigma \to f_0(500)$$\;\;$ & $\;\;\omega \to \omega(782) \,, \quad   [\sigma,\chi] \to [f_0(500),f_0(1370)]\;\;$\\ \hline
 chiral symmetry & always broken & dynamically broken by $\langle\sigma\rangle$\\ \hline
 parameters & \multicolumn{2}{c}{$\;\;$fitted to saturation properties at $B=0$$\;\;$}  \\ \hline
 magnetic field & \multicolumn{2}{c}{$\;\;$$\bar\psi i \gamma^\mu(\partial_\mu +iQA_\mu) \psi\;\;$ with 
$\;\;Q={\rm diag}(0,e)$, $\;\;A_\mu = (0,-yB,0,0)$$\;\;$}  \\ \hline
\end{tabular}
\caption{Comparison of the two models used here. In both cases we consider isospin-symmetric nuclear matter, the only difference between neutrons and protons 
being their electric charge. The magnetic field in the $z$ direction is introduced as a background gauge field, resulting in the fermion dispersions for the 
protons given in 
Eq.\ (\ref{epskn}). The parameters of both models are fitted to reproduce the density $n_0 = 0.153\,{\rm fm}^{-3}$, binding energy $E_{\rm bind} = -16.3 \, {\rm MeV}$, 
compression modulus $K = 250\,{\rm MeV}$, and effective mass $M_N = 0.8\,m_N$ with $m_N=939\,{\rm MeV}$, all at saturation and $B=0$. 
In the extended linear sigma model (eLSM), 
an additional scalar field $\chi$ is introduced which, via mixing with the $\sigma$, gives rise to two baryon masses, the nucleon and its chiral partner 
$N(1535)$, and to the two scalar resonances $f_0(500)$, $f_0(1370)$.}
\label{table0}
\end{table}

We work with two different phenomenological models, the Walecka model \cite{Walecka:1974qa} including scalar self-interactions \cite{Boguta:1977xi}, and an extended 
linear sigma model that includes the chiral partner of the nucleon via the "mirror assignment" \cite{Detar:1988kn,Gallas:2009qp,Gallas:2011qp,Heinz:2013hza}, 
both evaluated in the mean-field approximation. There are several reasons to use two different models. Firstly, we would like to check the model dependence of our results.
Secondly, the effect of magnetic catalysis is related to spontaneous breaking of chiral symmetry, and in this regard the two models are different: in the Walecka model, 
chiral symmetry is always broken by construction (the nucleon mass is a given parameter of the model), while the linear sigma model can account for chirally
broken and (approximately) restored phases (the nucleon mass is solely generated by spontaneous breaking of chiral symmetry). 
Therefore, magnetic catalysis can only be seen indirectly in the 
Walecka model, while the dependence of the chiral condensate on the magnetic field can be observed directly in the linear sigma model. And thirdly, we shall see that the 
presence of the chiral partner of the nucleon does make a difference because it has an effect on the nucleon mass in a magnetic field if the magnetic field is 
sufficiently large. We summarize the
most important properties and differences of the models in Table \ref{table0}. For more details and the Lagrangians of the models, see Ref.\ \cite{Haber:2014ula}. 

\begin{figure}
  \includegraphics[height=.28\textheight]{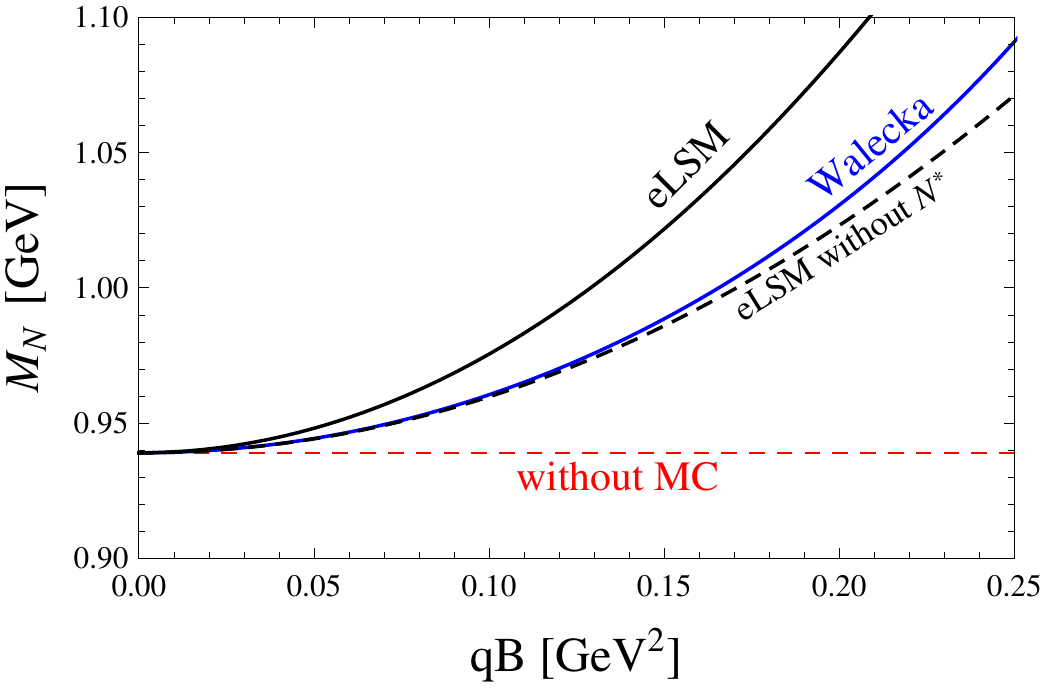}
  \caption{Vacuum mass of the nucleon in the presence of a background magnetic field, calculated in the Walecka and the extended linear sigma model. For comparison, 
we also show the result of the extended linear sigma model  
with the chiral partner of the nucleon $N^*$ removed by hand and the result without the sea contribution, 
which is responsible for magnetic catalysis (MC).}
\label{fig1}
\end{figure}

We restrict ourselves to zero temperature and start with the vacuum solution, i.e., we set $\mu=T=0$. In both models, the free energy is then given by the sea
contribution of the proton (\ref{sea}) and the tree-level potential of the mesons (since we neglect the $B$-independent sea contribution and the anomalous
magnetic moment of the nucleons, there is
no contribution of the neutron Dirac sea). We then extremize the free energy with respect to the meson 
condensates at a fixed, and thus irrelevant, value of the renormalization scale $\ell$. 
The result for the nucleon mass $M_N$ is shown in Fig.\ \ref{fig1}. 
We see that the mass increases monotonically with the magnetic field in both models,
in accordance with magnetic catalysis. The main difference of the models originates from the chiral partner of the nucleon which enhances the catalysing
effect and which is absent in the Walecka model. 

\section{Baryon onset and binding energy}

\begin{figure}
  \includegraphics[height=.28\textheight]{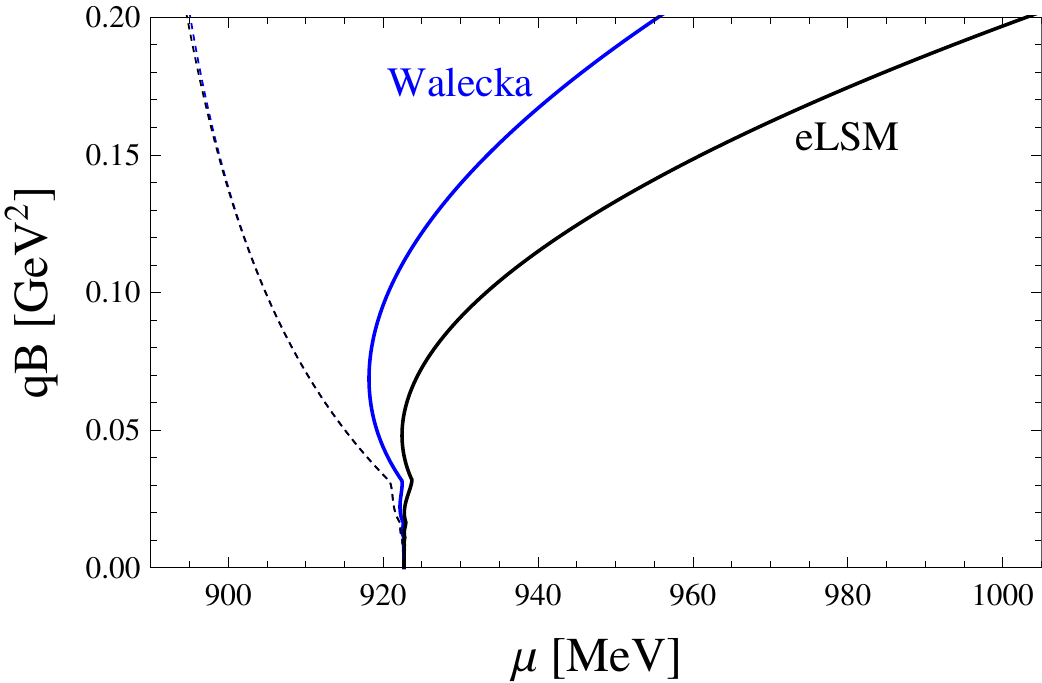}
  \caption{Zero-temperature phase diagram for the Walecka model, the extended linear sigma model, and both models without magnetic catalysis (dashed lines, which
are barely distinguishable). The curves are first-order phase transition lines that divide the vacuum (left) from nuclear matter (right).}
\label{fig2}
\end{figure}

\begin{figure}
  \includegraphics[height=.28\textheight]{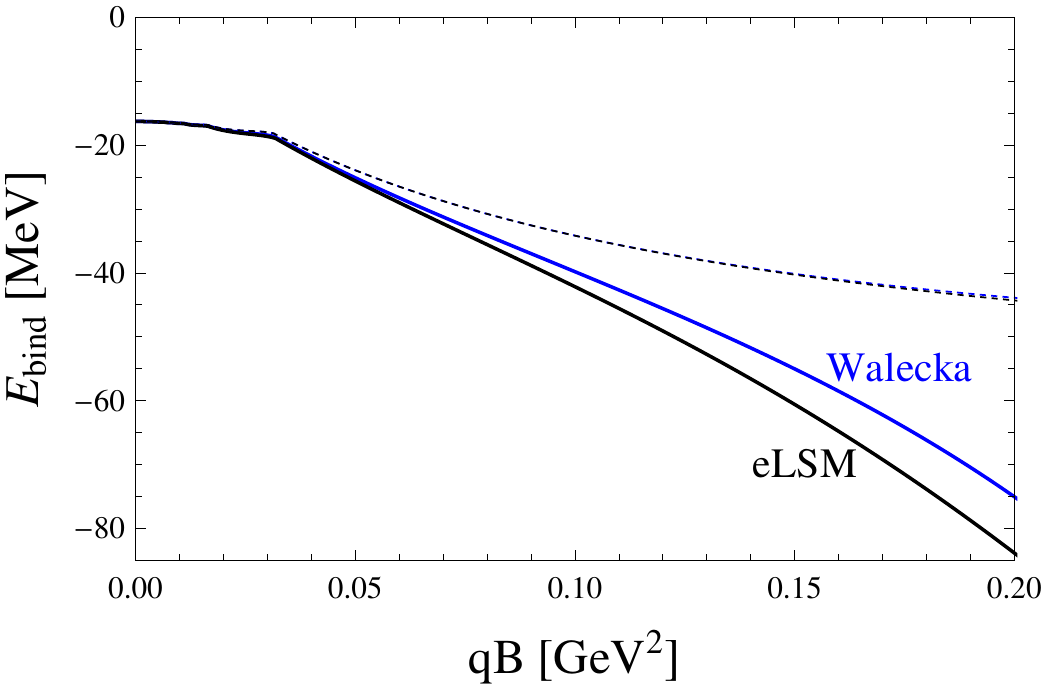}
  \caption{Binding energy along the baryon onset shown in Fig.\ \ref{fig2} for both models (solid lines) and without the Dirac sea contribution (dashed lines).}
\label{fig3}
\end{figure}

Next, we include the medium contribution of the free energy (\ref{Omega}), but continue to work at zero temperature. We focus on the baryon onset, i.e., now, together
with the extremization with respect to the condensates, we simultaneously require the free energies of the vacuum phase and the nuclear matter phase to be identical. 
This calculation yields the 
critical chemical potential $\mu_0$ for the baryon onset as a function of the magnetic field, with $\mu_0=922.7\, {\rm MeV}$ at vanishing magnetic field by construction.
The result is shown in Fig.\ \ref{fig2}. We can extract the binding energy from this result, $E_{\rm bind} = \mu_0 - M_N$, where $\mu_0$ is the $B$-dependent 
critical chemical potential, and $M_N$ the $B$-dependent vacuum mass, i.e., we separate the effect of the monotonic increase of the vacuum mass. At vanishing
magnetic field, $M_N=939\,{\rm MeV}$ and thus $E_{\rm bind} = -16.3\,{\rm MeV}$, again by construction. 
The binding energy as a function of the magnetic field is plotted in Fig.\ \ref{fig3}. 
Both plots show de Haas-van Alphen oscillations at small magnetic fields, $qB\lesssim 0.03 \,{\rm GeV}^2$, before all nucleons sit in the lowest Landau level $\nu=0$ at 
larger magnetic fields. The result for the binding energy shows that there are two counteracting effects on the critical chemical potential: 
the (modulus of the) binding energy increases and thus 
makes the creation of nuclear matter {\it less} costly, while the increasing vacuum mass due to magnetic catalysis has the opposite effect. 
At sufficiently large magnetic 
fields, magnetic catalysis dominates and the onset is shifted to larger chemical potentials. For comparison, we have plotted the onset curve {\it without} magnetic 
catalysis, i.e., without the $B$-dependent Dirac sea contribution (dashed curves). In this case, the critical chemical 
potential decreases monotonically (except for barely 
visible oscillations at small magnetic fields), and appears to saturate for $qB\to \infty$. The dashed curve in Fig.\ \ref{fig3} shows that the Dirac sea contribution is 
not only responsible for the increase in the vacuum mass, but also for an enhancement of the binding energy. 

The calculation shows different results for the two models. 
As already indicated in Fig.\ \ref{fig1}, the main difference comes from the chiral partner of the nucleon. We have checked that the onset and the binding 
energy of the two models are basically indistinguishable if we remove the chiral partner by hand from the extended linear sigma model. Of course,   
the chemical potentials considered here are too small to create a Fermi sea of the chiral partner. It is only the Dirac sea, i.e., the vacuum contribution, through which 
the chiral partner enters. In other words, the medium contribution is only relevant for $\mu> M_{N^*}$ (at zero temperature, this is a sharp limit), while the 
Dirac sea contribution is relevant for $qB \gtrsim M_{N^*}^2$ (this is {\it not} a sharp limit, and thus we see an effect). This observation shows that it is important 
to include other charged states, which we have ignored, into the calculation, for instance pions and hyperons.


\begin{theacknowledgments}
This work has been supported by the Austrian science foundation FWF under project no.~P26328-N27,  
and by the NewCompStar network, COST Action MP1304. 

\end{theacknowledgments}

\bibliographystyle{aipproc}   

\bibliography{references}

\end{document}